\documentstyle[11pt,epsfig,epsf,axocolor]{article}
\newcommand{\ba}{\begin{array}}
\newcommand{\ea}{\end{array}}
\newcommand{\bd}{\begin{displaymath}} 
\newcommand{\ed}{\end{displaymath}}
\newcommand{\be}{\begin{equation}}
\newcommand{\ee}{\end{equation}}
\newcommand{\bea}{\begin{eqnarray}}
\newcommand{\eea}{\end{eqnarray}}

\begin{document}
\begin{center}
{\Large\bf Heavy quark production via leptoquarks at a neutrino factory}
\\[15mm]
\vskip 15pt
{\sf Ashok Goyal $^{a, \,\!\!}$
\footnote{E-mail address: agoyal@ducos.ernet.in }},
{\sf Poonam Mehta $^{a, \,\!\!}$
\footnote{E-mail address: pmehta@physics.du.ac.in, mpoonam@mri.ernet.in }}
and
{\sf Sukanta Dutta $^{b, \,\!\!}$
\footnote{E-mail address: Sukanta.Dutta@cern.ch}}

\vskip 8pt
$^a${\em Department of Physics \& Astrophysics,
University of Delhi, Delhi 110 007, India }\\


$^b${\em Physics Department, S.G.T.B. Khalsa College, University of
Delhi, Delhi 110 007, India }\\

\end{center}
\vskip 5pt
\begin{abstract}
 
  The proposed neutrino factory (NF) based on a muon storage ring (MSR) 
  is an ideal place to look for heavy quark production  
  via neutral current (NC) and charged current (CC) interactions.
  In this article, we address the issue of contribution coming from mediating
  leptoquarks (LQ) in $\nu_\mu (\bar \nu_e)-{\rm N}$ scattering 
leading to the production of $b (\bar b)$ at a MSR 
and investigate the region where LQ interactions are significant
  in the near-site experiments.
\end{abstract}

\vskip 1cm
\setcounter{footnote}{0}
\begin{section}{Introduction}
It is widely believed that the proposed NF based on MSR capable of
supplying a well calibrated and intense beam of roughly
$\approx \,\, 10^{20} \,\, \nu_\mu ( \bar \nu_\mu) $ and
$\bar\nu_e (  \nu_e ) $ per year through 50 GeV muon  decays,
will open up unprecedented opportunity to reveal
the world of neutrino and to provide physical laboratory for testing
physics beyond the Standard Model (SM) \cite{storage_ring, hepph010}.
Recent strong indications of atmospheric neutrino oscillation 
($\nu_\mu \longrightarrow \nu_x$, where $x$ is not $e$) \cite{fukuda}
have rekindled the interest in accelerator
experiments that could study the same range of parameter space.
The solar neutrino deficit is interpreted either as matter
enhanced Mikhyev$-$Smirnov$-$Wolfenstein 
(MSW) oscillations \cite{solarmsw} or as vacuum
oscillations \cite{solarvac} that deplete the
original $\nu_e$'s, presumably in favour of $\nu_\mu$'s.
The role of a NF in determining masses and mixing angles
for $\nu_\mu \leftrightarrow \nu_\tau$ and 
$\bar\nu_e \leftrightarrow \bar\nu_\mu$ oscillations both at short
and long baseline experiments has been extensively discussed in the
literature. Investigation of physics beyond the SM through certain novel 
interactions in the neutrino sector, in particular the appearance
of $\tau$ and wrong sign $\mu$ signals in new physics
scenarios like supersymmetric (SUSY) theories with broken 
R-parity \cite{rp} and theories that allow LQ mediated 
lepton flavor violating (LFV) interactions \cite{lq} 
have been dealt with in our earlier works (\cite{rp},\cite{lq}).
With the same motivation to look for the role played by the 
non-standard interactions at a NF the production of heavy 
quarks through $\nu_\mu$-N scattering in an 
R-parity violating SUSY theory was investigated 
recently \cite{heavy} and it was shown that 
it is possible to have significant event rates for 
$b$ ($\bar b$) production via both NC and CC interactions. 
We should emphasize here that in SM the production of $b$ ($\bar b$) is
severely suppressed at tree level. 
Thus a considerable number of $b$ ($\bar b$) or an excess well 
above the SM rate at a NF would unequivocally imply the existence of 
non-standard physics in the neutrino sector. 
In contrast to SM, $b$ quark production via non-standard
$\nu-N$ scattering processes can take place
at the tree level itself via the CC interactions, 
$\nu_\mu \,\bar u \longrightarrow \mu^- \,\bar b$,
$\nu_\mu \,\bar c \longrightarrow \mu^- \,\bar b$
and $\bar\nu_e \, u \longrightarrow e^+ \, b$,
all of which are suppressed in the SM either due to the
Cabibbo$-$Kobayashi$-$Masakawa (CKM) matrix elements
$V_{ub}$ or due to the interaction of $\nu_\mu$
with sea quarks present inside the nucleon.
The corresponding NC processes 
$\nu_\mu \, d  \longrightarrow \nu_\mu \, b$
and $\bar\nu_e \, d  \longrightarrow \bar\nu_e\, b$
can occur only at one loop level in the SM.

\par In this context, it is worthwhile to consider theories with 
leptoquarks  which occur naturally in Grand Unified Theories,
Superstring inspired $E_6$ models and in Technicolor models
\cite{lqs} and study heavy flavour ($b,\bar b$) production
in scattering of neutrinos on a fixed isonucleon target
with LQ as mediators of the interaction.
In our earlier work, we have studied the contribution
of mediating lepton flavor violating 
LQ in $\nu_\mu\,(\bar\nu_e)$-N scattering leading to an enhanced 
production of $\tau$'s and wrong sign $\mu$'s at MSR and 
investigated the region where LQ interactions are significant 
in the near-site and short baseline experiments
and we found that {\it one can constrain LFV couplings
between the first and third generation, the bounds on which
are not generally available.}
With the same spirit in this present work,
we investigate the $b$ quark production in both NC and CC channels through
$\nu_\mu (\bar \nu_e) $-N scattering at the NF, mediated by scalar and 
vector leptoquarks. 
It is worth mentioning that we consider $\bar\nu_e$ 
beam also for production of $b,\bar b$ 
in both the NC and CC channels unlike reference \cite{heavy}.
For the present case since we are interested in new physics
effects alone and not the oscillation effects,
it is desirable to confine ourselves to near-site experiments
where the neutrino detectors are placed at a very short distance
(typically 40 m) from the storage ring.
Here we do not consider the LFV processes. 
The processes that we consider in this article  for the $b,\bar b$
production via NC and CC channels are :
\be
{\rm NC : ~} \nu_\mu \, d  \longrightarrow \nu_\mu \, b, \hskip 1 cm  \bar\nu_e \,
d  \longrightarrow \bar\nu_e\, b\label{eqn:nc}
\ee
\be
{\rm CC : ~} \nu_\mu \,\bar u \longrightarrow \mu^- \,\bar b,\hskip 1 cm  \bar\nu_e \,
u \longrightarrow e^+ \, b\label{eqn:cc}
\ee
The total number of $b,\bar b$ quark production events 
per year via either CC or NC interactions can
be written as
\begin{equation}
{\cal N}_{b,\,\bar b} = {\cal N}_n \int {
{{ d^2 \sigma^{\nu,{\bar \nu}}_{NC/CC}} \over {dx~dy} }}
\left[{d N_{\nu,\bar \nu} \over d E_{\nu_i,\bar \nu_i}}\right]
\,{\cal P}_{surv} (\nu_i(\bar \nu_i) \longrightarrow \nu_i(\bar \nu_i))
d E_{\nu_i(\bar \nu_i)}\,q(x)\,~dx~dy\label{eqn:n}
\end{equation}
\noindent where, ${\cal N}_n$ is the number of nucleons per kT
of the target material
{\footnote {${\cal N}_n = 6.023\, {\rm X}\, 
10^{32} ~{\rm for ~a ~target ~of ~mass ~1 ~kT}.$} }, 
$x$ and $y$ are the Bjorken scaling variables, 
q and q' are the quarks in the initial and final states, respectively and 
$q(x)$ is the quark distribution function.
The differential parton level cross-section can be expressed as 
\be
{{ d^2 \sigma^{\nu,{\bar \nu}}_{NC/CC}}
\over  {d x~ d y} }=  \left[{{ d^2 \sigma^{\nu,{\bar \nu}}_{NC/CC}}
\over  {d x'~ d y'} }\right] \times {\partial\left(x',\,y'\right)
\over\partial\left(x,\, y\right)} =
\left[{ \left\vert{\cal M}\left(\,x',\,y'\,\right)\right\vert ^2_{NC/CC}
\over {32 \pi \hat S} }\right] \label{eqn:xs}
\ee
where  $y'= - \,\hat t/\hat S = Q^2/ (2\, M \,E_\nu\, x')$, $x'$ is the
{\it slow rescaling} variable{\footnote {For production of a heavy quark from a light 
quark, the heavy quark mass modifies the scaling variable of the quark distribution.
$x'$ is the quark momentum fraction appropriate to absorb the virtual $W$ described by 
$\nu$ and $Q^2$.}} that arises due to the mass shell constraint of the heavy quark 
produced in the final state,
\be 
x'= {Q^2+m^2_Q\over 2\, M\,\nu}=x\, +\, {m^2_Q\over 2\, M\, E_\nu \,y}\hskip 1 cm
{\rm Therefore }\,\,\,\,\, \,
{\partial\left(x',\,y'\right)\over\partial\left(x,\, y\right)}=1.
\label{eqn:xyprime}
\ee
\noindent with $M$ being the nucleon mass, $E_\nu$ being the neutrino energy 
and $\nu=E_{\nu_l}-E_{l^-} \,\, (E_{\bar\nu_l}-E_{l^+}) $.  
$\hat S$ is the parton level CM energy and $\left[{d N_{\nu,\bar \nu}
\over d E_{\nu_i,\bar \nu_i}}\right]$ is the differential $\nu$
($\bar \nu$) flux.
The  survival probability of a particular neutrino flavour ($i$) is 
given by
${\cal P}_{surv} (\nu_{i}\rightarrow \nu_i) = $  
$1-{\cal P}_{osc} (\nu_{i}\rightarrow \nu_j)$ where j takes all possible 
values, $j = e,\mu,\tau$ but $j \neq i${\footnote
{For two flavour oscillation case,
$ {\cal P}_{osc} (\nu_{i}\rightarrow \nu_j)
= \sin^2 2\theta_m\,\, \sin^2\left[
1.27\, \Delta m^2[eV^2]\, {L[km]\over E_\nu[GeV]}\right]$,
where, $L$ is the baseline length, $E_\nu$ is the neutrino
energy, $\Delta m^2$ is the mass-squared difference between
the corresponding physical states, and $\theta_m$ is mixing angle
between flavours.}}.
\par The effective Lagrangian with the most general dimensionless,
$SU(3)_c$X$SU(2)_L$X$U(1)_Y$ invariant couplings of {\it scalar} and {\it
vector} LQ satisfying baryon ($B$) and lepton number ($L$) conservation
(suppressing colour, weak isospin and generation (flavour) indices )
is given \cite{lq_lag} by:
\bea
{\cal {L}} &=& {{\cal {L}}_{\vert F\vert =2}} + {{\cal {L}}_{\vert 
F\vert =0}}\,\,\,\,\, \,\,\,\, {\rm where} \nonumber\\
{{\cal {L}}_{\vert F\vert =2}} &=& \left[ g_{1L}\, \bar q^c _L\, i\, 
\tau_2 \,l_L +\, g_{1R}
\,\bar u^c_R \,e_R \right] \,S_1 +\,\tilde g_{1R}\, \bar d^c_R \,e_R 
\,\tilde S_1
+\,g_{3L}\,\bar q^c_L \,i \,\tau_2 \,{\vec \tau} \,l_L \,{\vec 
S}_3\nonumber \\
&+& \,\left[ g_{2L} \,\bar d^c _R \,\gamma^\mu \,l_L + \,g_{2R} \,\bar 
q^c_L\,
\,\gamma^\mu \,e_R \right] \,V_{2\mu} + \,\tilde g_{2L} \,\bar u^c_R 
\,\gamma^\mu \,l_L\,\tilde V_{2\mu} + \,{\rm \bf c.c.},\nonumber \\
{{\cal {L}}_{\vert F\vert=0}} &=& \left[ h_{2L} \,\bar u_R \,l_L +\, h_{2R}
\,\bar q _L\, i \,\tau_2 \,e_R \right] \,R_2 + \,\tilde h_{2L} \,\bar d 
_R \,l_L
\,\tilde R_2 + \,\tilde h_{1R} \,\bar u _R \,\gamma^\mu \,e_R \,\tilde 
U_{1\mu}
\nonumber \\
&+&\left[ h_{1L} \,\bar q _L \,\gamma^\mu \,l_L +
\,h_{1R} \,\bar d _R \,\gamma^\mu \,e_R \right] \,U_{1\mu}
+ h_{3L} \,\bar q _L \,{\vec \tau} \,\gamma^\mu \,l_L \,U_{3\mu} + {\rm 
\bf c.c.}
\eea
\noindent where  $q_L$, $l_L$ are the left-handed quarks
and lepton doublets and $e_R$, $d_R$, $u_R$ are the right-handed
charged leptons, down- and up-quark singlets respectively .
The Scalar (i.e. $S_1$, $\tilde S_1$, {\bf $S_3$}) and
Vector (i.e. $V_2$, $\tilde V_2$) LQ carry fermion number
${\rm F}=3{\rm B}+{\rm L}=-2$,
while the Scalar (i.e. $R_2$, $\tilde R_2$ ) and Vector (i.e. $U_1$,
$\tilde U_1$, {\bf $U_3$}) LQ have ${\rm F}=0$.
\par Numerous phenomenological studies have been made in order to 
derive bounds and put stringent constraints on LQ couplings particularly 
from low energy FCNC processes
\cite{davidson} that are generated by scalar and
vector LQ interactions.
Direct experimental searches for leptoquarks have also been
carried out at the e-p collider and bounds obtained
\cite{davidson, lqbounds} and in particular bounds obtained
from $B$ meson decays $\left( B\longrightarrow l^+ l^- X, ~
{\rm where}~~ l^+ l^- = \mu^+\mu^-,~ e^+ e^- \right)$
and also bounds derived from meson-antimeson ($B\bar B$) mixing
would have direct bearing on the processes considered here.
This is because low energy limit puts stringent bound on
effective four-fermion interactions involving two leptons and two 
quarks and since at the NF the centre of mass energy
in collision is low enough, we can consider the neutrino-quark
interaction as an effective four-fermion interaction. 
The bounds on effective couplings used in this paper are the 
LQ couplings over mass squared of the LQ and are derived
on the assumption that individual leptoquark coupling contribution to
the branching ratio does not exceed the experimental upper limits and 
in the branching ratios only one leptoquark coupling is considered by
switching off all the other couplings. All couplings are considered 
to be real and combinations of left and right chirality
coupling are not considered.
This article is outlined as follows. 
We discuss the $b (\bar b)$ production 
through $\nu(\bar \nu)$-N 
interactions via NC and CC channels in section 2 and 3 respectively 
and give the plots of event rate versus muon beam energy. 
In section 4 we outline the conclusions drawn from our 
results.

\end{section}

\begin{section}{${\rm {b (\bar b)}}$ Production Via  NC Processes:}

\par Let us first consider the possible NC processes
that can lead to $b/\bar b$ in the final state.
There is no SM tree level process in the NC channel as 
NC processes leading to $b/\bar b$ can only occur at one loop level in 
the SM.
However, there can be two possible non-standard tree level
NC processes that can lead to the production of $b/\bar b$
in the final state, due to the presence of both
$\nu$ and $\bar \nu$ of different flavors from $\mu$ decay,
{\it viz $\mu^- \longrightarrow e^- \nu_\mu {\bar {\nu_e}}$}
\begin{enumerate}
\item    $\nu_\mu + d \longrightarrow \nu_\mu + b$
\item    $\bar \nu_e + d \longrightarrow \bar \nu_e + b$
\end{enumerate}
\noindent For the two NC processes mentioned above,
we have both s- and u-channel diagrams arising from the relevant
interaction terms in the effective LQ lagrangian.
For the first process, $\nu_\mu + d \longrightarrow \nu_\mu + b$ (
shown in figure 1 ), there are two possible u-channel diagrams
mediated by LQs $( {\tilde R}^\dag, {U}^\dag )$
carrying $\vert {\rm F} \vert = 0$ and charge $=1/3$,
while there are three possible s-channel diagrams
that are mediated by LQs $( {S}^\dag, {V}^\dag )$
carrying $\vert {\rm F} \vert = 2$ and charge $=-1/3$.
For the second process, $\bar \nu_e \,+\, d
\longrightarrow \bar \nu_e \,+\, b$ ( shown in figure 2 ),
the two possible s-channel diagrams are
mediated by LQs $( \tilde R, U )$ carrying $\vert {\rm F} \vert = 0$
and charge $=-1/3$, while the three possible u-channel diagrams are
mediated by LQs
$( S, V )$ carrying $\vert {\rm F} \vert = 2$ and charge $=1/3$.

\begin{figure}[h]
\begin{picture}(280,100)(0,0)
\vspace*{- 1in}
\ArrowLine(60,-5)(100,30){psRed}
\Text(60,-16)[r]{$\nu_\mu$}
\ArrowLine(100,30)(57,65){psRed}
\Text(60,80)[r]{${b}$}
\DashLine(100,30)(170,30){4}{psBlue}
\Text(135,40)[b]{$\tilde R,\,U$}
\ArrowLine(170,30)(200,70){psRed}
\Text(202,80)[b]{$\nu_\mu$}
\ArrowLine(210,-3)(170,30){psRed}
\Text(202,-16)[b]{$d$}
\Text(135,-15)[b]{(a)}

\ArrowLine(265,-5)(305,30){psRed}
\Text(265,-16)[r]{$\nu_\mu$}
\ArrowLine(262,65)(305,30){psRed}
\Text(265,80)[r]{$d$}
\DashLine(305,30)(375,30){4}{psBlue}
\Text(340,40)[b]{$S,\,V$}
\ArrowLine(375,30)(405,70){psRed}
\Text(402,76)[b]{$\nu_\mu$}
\ArrowLine(375,30)(415,-3){psRed}
\Text(407,-16)[b]{$b$}
\Text(330,-15)[b]{(b)}
\end{picture}
\vskip .3in
\caption{{\em  b production via NC process ($\nu_\mu + d
\longrightarrow \nu_\mu + b$) from scalar \& vector LQ: (a) u-channel
process corresponding to $\vert {\rm F}\vert=0$ LQ and
(b) s-channel process corresponding to $\vert{\rm }F\vert=2$ LQ.}}
\label{fig:dia1}
\end{figure}
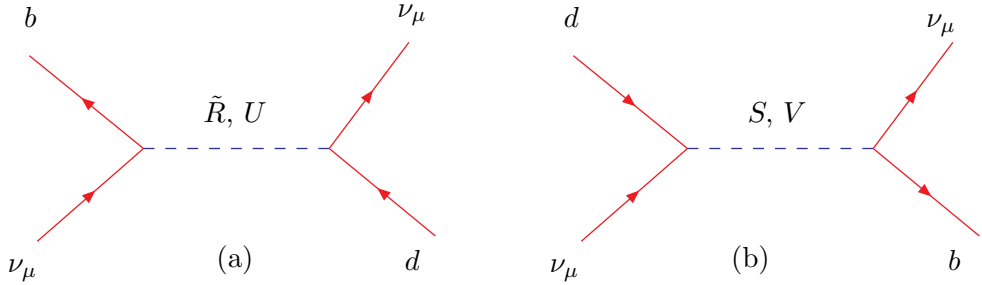

We first consider the production of ``b'' from $\nu_\mu$
(obtained from $\mu^-$ decay)
interactions with nucleon via NC u-channel
processes for $|F|=0$ case ( figure  \ref{fig:dia1}(a) )
and NC s-channel processes for $|F|=2$ case  ( figure  \ref{fig:dia1}(b) ).
There are in all two diagrams contributing to production of b via
($\nu_\mu + d \longrightarrow \nu_\mu + b$)
in the u-channel (figure \ref{fig:dia1}(a)),
one mediated by the charge $=1/3$, scalar LQ
$( {\tilde {R_2^{-1/2}}}^\dag )$ carrying $T_3=-1/2$
and the other one by a vector LQ $( {U_{3\mu}^-}^\dag )$
with $T_3=-1$, where $T_3$ is the weak isospin.
The matrix element squared for 2 diagrams contributing to the
u-channel NC process is
\bea
{ \left\vert{\cal M}_{LQ}^{u-chann}(\nu_\mu d \longrightarrow \nu_\mu b)
\right\vert ^2 } =
\left[ \hat u (\hat u - m_b ^2)\right]
\Biggl[{ \left\vert {\tilde {h_{2L}}} \, {\tilde {h_{2L}}}\right\vert ^2 
\over (\hat u - M_{ \tilde {R _{2} ^{-1/2} }}^2)^2 }\Biggr]
~+~ \left[4 \hat s (\hat s - m_b ^2)\right]
\Biggl[{\left\vert {\sqrt 2} h_{3L}\, {\sqrt 2} h_{3L} \right\vert ^2
\over (\hat u - M_{{U_{3\mu}^-}}^2)^2} \Biggr] \nonumber \\
\eea
\\
\noindent where, the Mandelstam variables at the parton level are given by
${\hat s}=(p_{\nu_\mu}+p_d)^2$, ${\hat t}=(p_{\nu_\mu}(initial)-
p_{\nu_\mu}(final))^2$ and
${\hat u}=(p_{\nu_\mu}-p_b)^2$, with $p_i$ denoting the
four momentum of the $i^{th}$ particle.

\par In the s-channel, two diagrams are mediated by charge $=-1/3$,
scalar LQs $( {S_1}^\dag,{S_3^0}^\dag )$
with $T_3=0$, while one is mediated by a
vector LQ $( {V_{2\mu}^{-1/2}}^\dag )$ with
$T_3=-1/2$ ( figure \ref{fig:dia1}(b) ).
The matrix element squared for all the 3 diagrams contributing to the NC
s-channel process is
\bea
{ \left\vert{\cal M}_{LQ}^{s-chann}(\nu_\mu d \longrightarrow \nu_\mu b)
\right\vert ^2 } =
\left[\hat s (\hat s - m_b ^2)\right] \Biggl[ { \left\vert g_{1L} \, g_{1L}
\right\vert ^2 \over (\hat s - M_{S_1}^2)^2} ~+~
{ \left\vert g_{3L} \, g_{3L} \right\vert^2 \over
(\hat s - M_{S_3^0}^2)^2}  \nonumber \\
~+~ 2 {\left\vert g_{1L} \, g_{3L}\right\vert^2 \over (\hat s - M_{S_1}^2) 
(\hat s - M_{S_{3}^0}^2) }\Biggr]
~+~ \left[ 4\hat u (\hat u - m_b ^2)\right]
\Biggl[{ \left\vert g_{2L}\,  g_{2L}
\right\vert^2 \over (\hat s - M_{ V _{2\mu} ^{-1/2}}^2)^2 } \Biggr]
\eea
\\

\begin{figure}[h]
\begin{picture}(280,100)(0,0)
\vspace*{- 1in}
\ArrowLine(60,-5)(100,30){psRed}
\Text(60,-16)[r]{${\bar {\nu_e}}$}
\ArrowLine(57,65)(100,30){psRed}
\Text(60,80)[r]{${d}$}
\DashLine(100,30)(170,30){4}{psBlue}
\Text(135,40)[b]{$\tilde R,\,U$}
\ArrowLine(170,30)(200,70){psRed}
\Text(202,80)[b]{${\bar {\nu_e}}$}
\ArrowLine(170,30)(210,-3){psRed}
\Text(202,-16)[b]{$b$}
\Text(135,-15)[b]{(a)}

\ArrowLine(265,-5)(305,30){psRed}
\Text(265,-16)[r]{${\bar {\nu_e}}$}
\ArrowLine(305,30)(262,65){psRed}
\Text(265,80)[r]{$b$}
\DashLine(305,30)(375,30){4}{psBlue}
\Text(340,40)[b]{$S,\,V$}
\ArrowLine(375,30)(405,70){psRed}
\Text(402,76)[b]{${\bar {\nu_e}}$}
\ArrowLine(415,-3)(375,30){psRed}
\Text(407,-16)[b]{$d$}
\Text(330,-15)[b]{(b)}
\end{picture}
\vskip .3in
\caption{{\em  b production via NC process
($\bar \nu_e \,+\, d \longrightarrow \bar \nu_e \,+\, b$) from scalar
\& vector LQ: (a) s-channel process
corresponding to $\vert {\rm F}\vert=0$ LQ and
(b) u-channel process corresponding to $\vert{\rm }F\vert =2$ LQ.}}
\label{fig:dia2}
\end{figure}
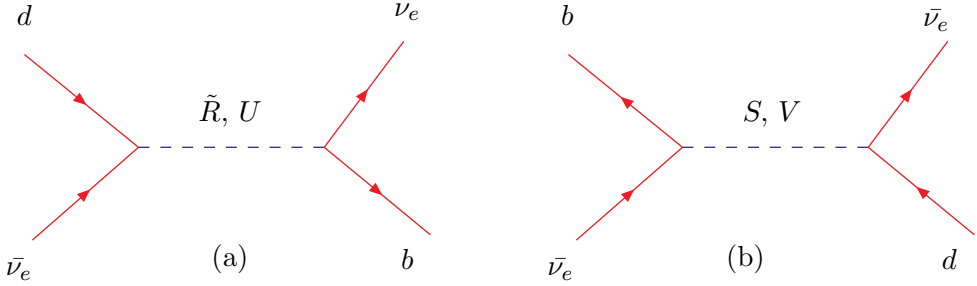

Next we consider the production of ``b'' from $\bar {\nu_e}$
(also obtained from the $\mu^-$ decay) through
interactions with nucleon via NC s-channel process
for $|F|=0$ case ( figure  \ref{fig:dia2}(a) )
and NC u-channel process for $|F|=2$ case  ( figure  \ref{fig:dia2}(b) ).
There are in all two diagrams contributing to production of b via
($\bar {\nu_e} \,+\, d \longrightarrow \bar {\nu_e} \,+\, b$)
in the s-channel ( figure \ref{fig:dia2}(a) ),
one mediated by the charge $=-1/3$, scalar LQ $( \tilde R_2^{-1/2} )$ with
$T_3=-1/2$ and while the other one by a vector LQ $( U_{3\mu}^- )$
with $T_3=-1$.
The matrix element squared for the 2 diagrams contributing to the
NC s-channel process is
\bea
{ \left\vert{\cal M}_{LQ}^{s-chann}(\bar {\nu_e} d
\longrightarrow \bar {\nu_e} b)\right\vert ^2 } =
\left[ \hat s (\hat s - m_b ^2)\right]
\Biggl[{ \left\vert {\tilde {h_{2L}}} \, {\tilde {h_{2L}}}
\right\vert ^2 \over (\hat s - M_{ \tilde {R _{2} ^{-1/2} }}^2)^2 }\Biggr]
~+~ \left[4 \hat u (\hat u - m_b ^2)\right]
\Biggl[{\left\vert {\sqrt 2} h_{3L}\, {\sqrt 2} h_{3L} \right\vert ^2
\over (\hat s - M_{{U_{3\mu}^-}}^2)^2} \Biggr] \nonumber \\
\eea
\\
\noindent where, the Mandelstam variables at the parton level are given by
${\hat s}=(p_{\bar {\nu_e}}+p_d)^2$,
${\hat t}=(p_{\bar {\nu_e}}(initial)-p_{\bar {\nu_e}}(final))^2$ and
${\hat u}=(p_{\bar {\nu_e}}-p_b)^2$.

\par In the u-channel, two diagrams are mediated by the charge $=1/3$,
scalar LQs $( S_1,S_3^0 )$ with $T_3=0$ and
one is mediated by a vector LQ $( V_{2\mu}^{-1/2} )$ with $T_3=-1/2$
( figure \ref{fig:dia2}(b) ).
The matrix element squared for all 3 diagrams contributing to the NC
u-channel process is
\bea
{ \left\vert{\cal M}_{LQ}^{u-chann}(\bar {\nu_e} d \longrightarrow
\bar {\nu_e} b)\right\vert ^2 } =
\left[\hat u (\hat u - m_b ^2)\right] \Biggl[ { \left\vert g_{1L} \, g_{1L}
\right\vert ^2 \over (\hat u - M_{S_1}^2)^2} ~+~
{ \left\vert g_{3L} \, g_{3L} \right\vert^2 \over
(\hat u - M_{S_3^0}^2)^2}  \nonumber \\
~+~ 2 {\left\vert g_{1L} \, g_{3L}\right\vert^2 \over (\hat u - M_{S_1}^2) 
(\hat u - M_{S_{3}^0}^2) }\Biggr]
~+~ \left[ 4\hat s (\hat s - m_b ^2)\right]
\Biggl[{ \left\vert g_{2L}\,  g_{2L}
\right\vert^2 \over (\hat u - M_{ V _{2\mu} ^{-1/2}}^2)^2 } \Biggr]
\eea
\\
\begin{figure}[h]
\vskip -4cm
\centerline{
\epsfxsize=20 cm\epsfysize=24cm\epsfbox{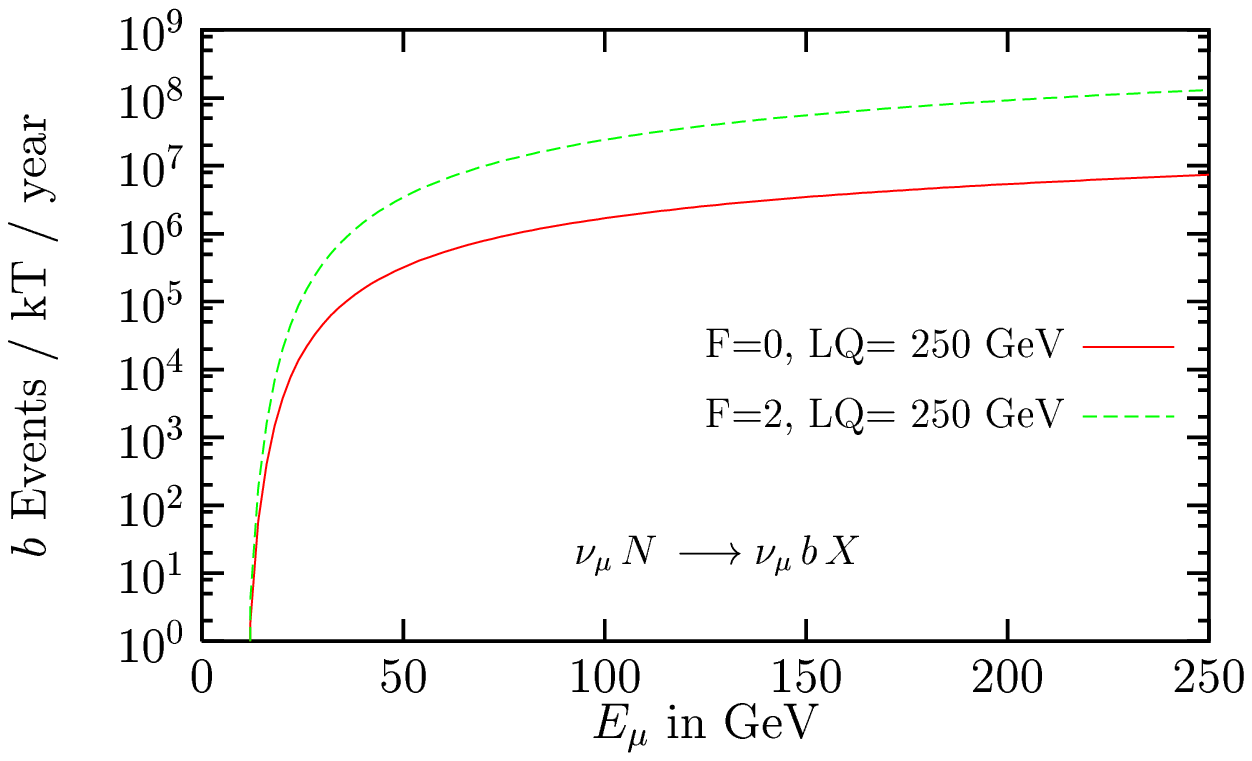}
}
\vskip -13cm

\caption{{\em Variation of $ b$-events ( from LQ ) for a 1kT detector and
LQ mass 250 GeV with muon beam energy for a baseline length 40 meters
and sample detector area 0.025 m$^2$}}
\label{fig:fig3}
\end{figure}
\begin{figure}[h]
\vskip -4cm
\centerline{
\epsfxsize=20 cm\epsfysize=24cm\epsfbox{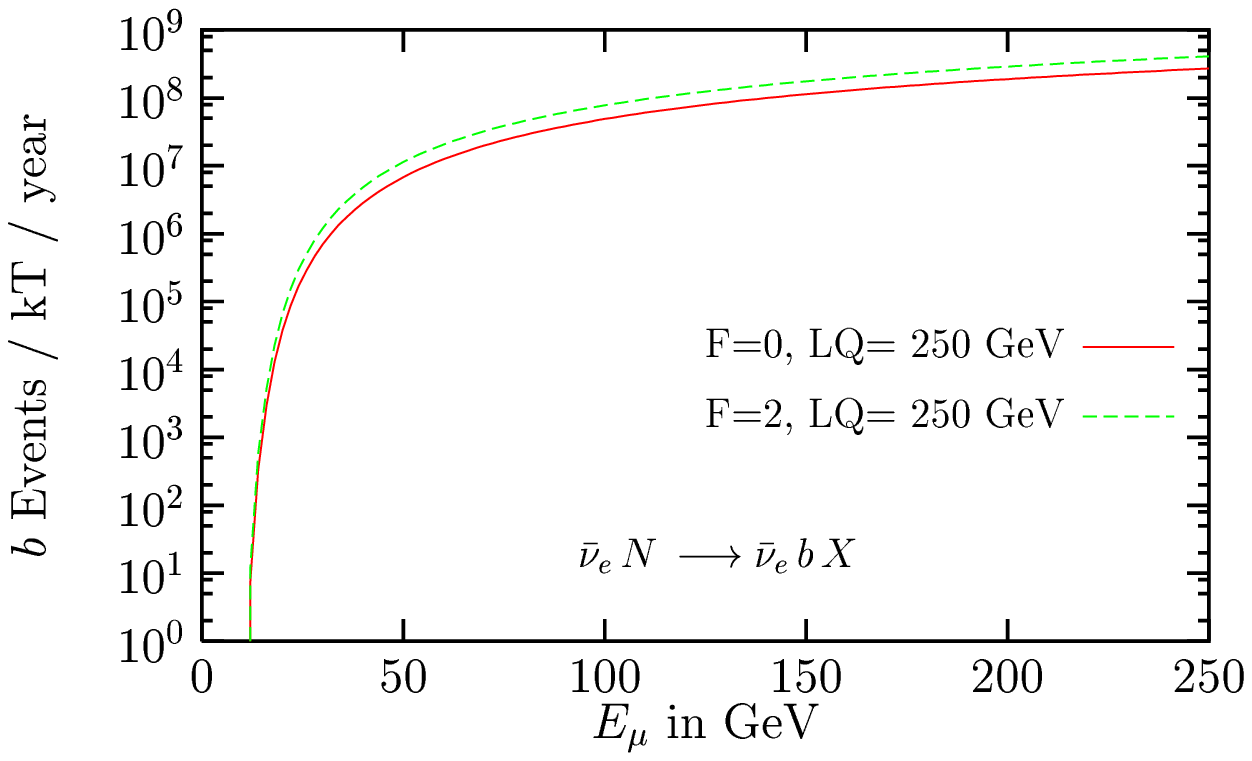}
}
\vskip -13cm

\caption{{\em Variation of $ b$-events ( from LQ ) for a 1kT detector and
LQ mass 250 GeV with muon beam energy for a baseline length 40 meters
and sample detector area 0.025 m$^2$}}
\label{fig:fig4}
\end{figure}
Having said all about the relevant NC diagrams leading to b-production,
we now focus on the details that we use in order to compute
the number of events for $b/\bar b$ via NC channel and demonstrate their
behaviour as a function of muon energy ranging from 0 upto 250 ${\rm GeV}$.
We consider the contribution from LQ carrying different fermion numbers
separately, which essentially
means that {\it either all the h's or all the g's}
contributing to a given process, are non-zero at a time.
For simplicity, we take the masses of scalar and vector
LQs for both ${\rm F}=0$ and $|{\rm F}|=2$ to be equal (~=~250 ${\rm 
GeV}$~).
As in our earlier works (~\cite{rp},~\cite{lq}~),
we have used CTEQ4LQ Parton Distribution Functions
\cite{cteq} in order to compute the events. There is however significant suppression in phase space due to the production of massive \lq $b$\rq \, quark. In our calculation we have not imposed any event selection cuts. Using events selection cuts for detail analysis as given in reference \cite{hepph010} will further scale down  the contribution.
We have considered 
a detector with a sample area of 0.025 $m^2$ \cite{detector} and placed at 40 m from the storage ring.
Regarding the bounds on LQ couplings, we have used model independent
constraints on the couplings to $b$ quarks of $B$ and $L$ conserving LQs
as discussed in \cite{davidson} where it is shown that one
can constrain the generation dependent LQ couplings to $b$ quarks
from the upper bounds on the flavour-changing decays
$B\longrightarrow l^+ l^- X$ (where $l^+ l^- = \mu^+ \mu^-, e^+ e^-$),
the CKM matrix element $V_{ub}$ and from meson\,-\,antimeson
($B\bar B$) mixing and obtain some of the best bounds for the processes of our interest.
All the bounds on couplings that we have used for calculation of event rates are listed in table
\ref{table1}. Since the bounds on the couplings $h_{2L} ~\&~g_{1R}$
are not available from reference \cite{davidson}, we take them to
be the same as bounds on couplings $h_{2R} ~\&~ g_{1L}$ (which
are the opposite chirality counterparts of $h_{2L} ~\&~g_{1R}$
respectively). 
We make some simplifying assumptions like the product of couplings of
different chirality is obtained from the squares of the
couplings of individual chiralities.
We extract bounds relevant to $(\nu_\mu d)(\nu_\mu b)$ vertex
from the bounds for (21)(23) generation of quark-lepton pair, while for the
vertex $(\bar \nu_e d)(\bar \nu_e b)$, we use
the bounds for the (11)(13) generation indices
relevant to the process. These bounds are 
derived from semileptonic inclusive B decays. 
The latest bounds coming from BABAR and BELLE experiments 
\cite{baba_belle} however are not relevant for the processes 
considered here except for the bound on $V_{ub}$ which does not 
make any significant change in the couplings.   
        In figure \ref{fig:fig3} and \ref{fig:fig4}, 
we have plotted the b-quark production rate as a function of muon beam 
energy for $\nu_\mu-{\rm N}$ and ${\bar \nu}_e-{\rm N}$ scattering processes 
respectively. 
\begin{center}
\begin{table}[htb]
\begin{tabular}{|p{.5in}|p{.4in}|p{.4in}|p{.4in}|p{.4in}|p{.4in}
|p{.4in}|p{.4in}|p{.4in}|p{.4in}|p{.4in}|p{.4in}|}\hline
\hline
(lq)(lq)& $h_{1L}$ &  $h_{1R}$ & $ h_{2L}$ & $ h_{2R}$ &
$ h_{3L}$ & $g_{1L}$ &  $g_{1R}$& $g_{2L}$ &$g_{2R}$ &  $g_{3L}$\\\hline 
\hline
(11)(13)&
.002 & .003&--  & .006 &  .002 & .004 &-- &.003 &.003 & .004\\\hline \hline
(21)(23)& .0004& .0004&--& .0008& .0004 &.004 &-- 
&.0004&.0004&.0004\\\hline
\hline
\end{tabular}

\caption{The best bounds on all relevant products of couplings
(from B decays and $B\bar B$ mixing) taken from table 15 of the
reference \cite{davidson} by S. Davidson {\it et al.}
).
All the bounds are multiplied by $(m_{LQ}/[100~ {\rm GeV}])^2$.}
\label{table1}
\end{table}
\end{center}

\end{section} 

\begin{section}{${\rm {b (\bar b)}}$ Production Via CC  Processes:}

\begin{figure}[h]
\begin{picture}(280,100)(0,0)
\vspace*{- 1in}
\ArrowLine(60,-5)(100,30){psRed}
\Text(60,-16)[r]{${\nu_\mu}$}
\ArrowLine(57,65)(100,30){psRed}
\Text(60,80)[r]{${\bar u}$}
\DashLine(100,30)(170,30){4}{psBlue}
\Text(135,40)[b]{$R,\,U$}
\ArrowLine(170,30)(200,70){psRed}
\Text(202,80)[b]{${\mu^-}$}
\ArrowLine(170,30)(210,-3){psRed}
\Text(202,-16)[b]{${\bar b}$}
\Text(135,-15)[b]{(a)}

\ArrowLine(265,-5)(305,30){psRed}
\Text(265,-16)[r]{${\nu_\mu}$}
\ArrowLine(305,30)(262,65){psRed}
\Text(265,80)[r]{${\bar b}$}
\DashLine(305,30)(375,30){4}{psBlue}
\Text(340,40)[b]{$S,\,V$}
\ArrowLine(375,30)(405,70){psRed}
\Text(402,76)[b]{${\mu^-}$}
\ArrowLine(415,-3)(375,30){psRed}
\Text(407,-16)[b]{${\bar u}$}
\Text(330,-15)[b]{(b)}
\end{picture}
\vskip .3in
\caption{{\em b production via CC process ($\nu_\mu + \bar u
\longrightarrow \mu^- + \bar b$) from scalar \& vector LQ: (a)
s-channel diagram corresponding
to $\vert {\rm F}\vert=0$ LQ and (b) u-channel diagram corresponding
to $\vert {\rm F}\vert=2$ LQ.}}
\label{fig:dia5}
\end{figure}
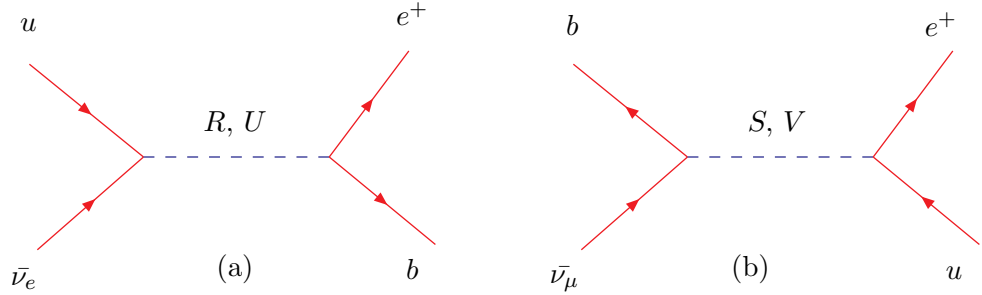
\begin{figure}[h]
\begin{picture}(280,100)(0,0)
\vspace*{- 1in}
\ArrowLine(60,-5)(100,30){psRed}
\Text(60,-16)[r]{${\bar {\nu_e}}$}
\ArrowLine(57,65)(100,30){psRed}
\Text(60,80)[r]{${u}$}
\DashLine(100,30)(170,30){4}{psBlue}
\Text(135,40)[b]{$R,\,U$}
\ArrowLine(170,30)(200,70){psRed}
\Text(202,80)[b]{${e^+}$}
\ArrowLine(170,30)(210,-3){psRed}
\Text(202,-16)[b]{${b}$}
\Text(135,-15)[b]{(a)}

\ArrowLine(265,-5)(305,30){psRed}
\Text(265,-16)[r]{${\bar {\nu_\mu}}$}
\ArrowLine(305,30)(262,65){psRed}
\Text(265,80)[r]{${b}$}
\DashLine(305,30)(375,30){4}{psBlue}
\Text(340,40)[b]{$S,\,V$}
\ArrowLine(375,30)(405,70){psRed}
\Text(402,76)[b]{${e^+}$}
\ArrowLine(415,-3)(375,30){psRed}
\Text(407,-16)[b]{${u}$}
\Text(330,-15)[b]{(b)}
\end{picture}
\vskip .3in
\caption{{\em  b production via CC process ($\bar \nu_e + u
\longrightarrow e^+ + b$) from scalar \& vector LQ: (a) s-channel diagram
corresponding to $\vert {\rm F}\vert=0$ LQ and
(b) u-channel diagram corresponding to $\vert {\rm F}\vert=2$ LQ.}}
\label{fig:dia6}
\end{figure}

As discussed above the production of $\bar b$ or $b$ in the 
final state through CC interaction can also occur in the SM at 
the tree level in contrast 
to the NC case where SM contributes only at the one loop level. 
The SM cross-sections for the CC processes 
$\nu_\mu + \bar u \longrightarrow \mu^- + \bar b$
and $\bar \nu_e + u \longrightarrow e^+ + b$ are given by 
\bea
{ d^2 \sigma \over  d x~ d y } \left(\nu_\mu\, N \longrightarrow
\mu^-\, \bar b\, X\,\right)&=&{G^2_F\,  S\over \pi } \left({M_W^2
\over M_W^2 + Q^2}\right)^2\,\left( x'-x'\,y'-{m^2_b\over  S}\right)
\left(1-y'\right) \bar u(x') \left\vert V_{ub}\right\vert^2\nonumber\\
{ d^2 \sigma \over  d x~ d y } \left(\bar\nu_e\, N \longrightarrow
e^+ \, b\, X\,\right)&=&{G^2_F\,  S\over \pi } \left({M_W^2\over M_W^2 + 
Q^2}
\right)^2\,\left( x'-x'\,y'-{m^2_b\over  S}\right) \left(1-y'\right)  u(x')
\left\vert V_{ub}\right\vert^2\nonumber\\
&& \label{eqn:sm}
\eea
Here we have the advantage of having the SM rates as benchmark against which to 
the compare the rates obtained via  LQ. ${\bar u}(x')$ and $u(x')$ are the  
distribution functions of up-type antiquark and quark respectively. 

\noindent For the CC processes mentioned above,
we can have both s- and u-channel diagrams arising from the relevant
interaction terms in the effective LQ Lagrangian, as for the case of NC
processes. For the first process,
$\nu_\mu + \bar u \longrightarrow \mu^- + \bar b$ ( as shown in figure 3 ),
there are 4 possible s-channel diagrams mediated by LQs $( {R}^\dag,
{U}^\dag )$ carrying $\vert {\rm F} \vert = 0$ and charge $=-2/3$. Also
there are 4 possible u-channel diagrams that are mediated by LQs $( 
{S}^\dag,{V}^\dag )$ carrying $\vert {\rm F} \vert = 2$ and charge $=-1/3$.
For the second process,
$\bar \nu_e + u \longrightarrow e^+ + b$ ( as shown in figure 4 ),
the 4 possible s-channel diagrams are mediated by LQs $( R, U )$
carrying $\vert {\rm F} \vert = 0$ and charge $=2/3$,
while the 4 possible u-channel diagrams are mediated by LQs $( S, V )$
carrying $\vert {\rm F} \vert = 2$ and charge $=1/3$, respectively.
We first consider the production of ``$\bar b$'' from $\nu_\mu$
(obtained from $\mu^-$ decay) through
interactions with nucleon via CC s-channel process
for the $|F|=0$ case  ( figure 5(a) )
and CC u-channel process for the $|F|=2$ case  ( figure 5(b) ).

\begin{figure}[h]
\vskip -4cm
\centerline{
\epsfxsize=20 cm\epsfysize=24cm\epsfbox{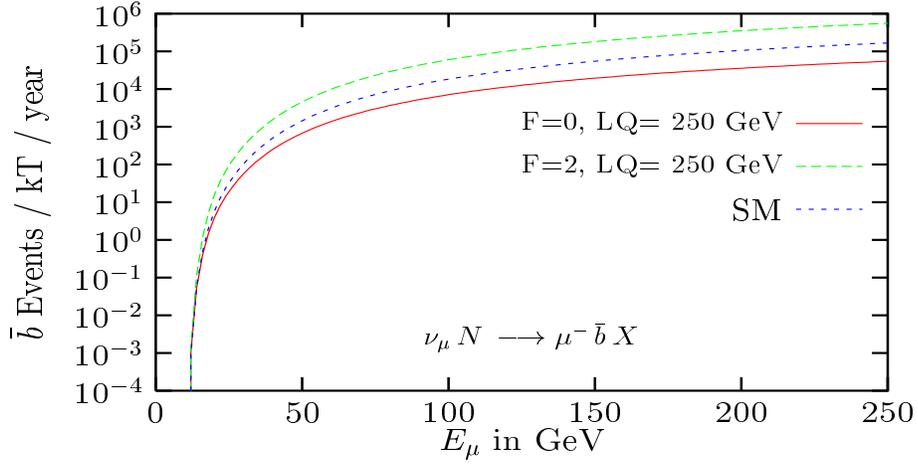}
}
\vskip -13cm

\caption{{\em Variation of $\bar b$-events ( from SM and LQ ) for a 1kT
detector and
LQ mass 250 GeV with muon beam energy for a baseline length 40 meters
and sample detector area 0.025 m$^2$}}
\label{fig:fig7}
\end{figure}

There are in all 4 diagrams contributing to production of ${\bar b}$ via
($\nu_\mu + {\bar u} \longrightarrow \mu^- + {\bar b} $)
in the s-channel ( figure 5(a) ),
one mediated by the charge $=-2/3$,
scalar LQ $( {R_2^{-1/2}}^\dag )$ with $T_3=-1/2$ and
the other three by vector LQs $( {U_{1\mu}}^\dag, {U_{1\mu}}^\dag,
{U_{3\mu}^0}^\dag )$ with $T_3=-1$.
The matrix element squared for all 4 diagrams contributing to the
CC s-channel process is
\bea
{ \left\vert{\cal M}_{LQ}^{s-chann}(\nu_\mu {\bar u}
\longrightarrow \mu^- {\bar b})\right\vert ^2 } =
\left[ \hat s (\hat s - m_b ^2)\right]
\Biggl[{ \left\vert { {h_{2L}}} \, { {h_{2R}}}\right\vert ^2
\over (\hat s - M_{ {R _{2} ^{-1/2} }}^2)^2 }\Biggr]
~+~
\left[4 \hat u (\hat u - m_b ^2)\right]
\Biggl[{\left\vert h_{1L}\, h_{1L} \right\vert ^2
\over (\hat s - M_{{U_{1\mu}}}^2)^2}
\nonumber \\
~+~
{\left\vert h_{3L}\, h_{3L} \right\vert ^2
\over (\hat s - M_{{U_{3\mu}^0}}^2)^2}
~-~
2\,{\left\vert h_{1L}\, h_{3L} \right\vert ^2
\over
(\hat s - M_{{U_{1\mu}}}^2) \, (\hat s - M_{{U_{3\mu}^0}}^2)}
\Biggr]
~+~
\left[4 (\hat s + \hat u)(\hat s + \hat u - m_b ^2)\right]
\Biggl[{\left\vert h_{1L}\, h_{1R} \right\vert ^2
\over (\hat s - M_{{U_{1\mu}}}^2)^2}
\Biggr]
\eea
\\
\noindent where, the Mandelstam variables at the parton level are given by
${\hat s}=(p_{\nu_\mu}+p_{\bar u})^2$,
${\hat t}=(p_{\nu_\mu}-p_{\mu^-})^2$ and
${\hat u}=(p_{\nu_\mu}-p_{\bar b})^2$.

\par In the u-channel, 3 diagrams are mediated by the charge $=-1/3$,
scalar LQs $( {S_1}^\dag, {S_1}^\dag, {S_3^0}^\dag )$
with $T_3=0$ and one is mediated by a vector LQ $( 
{V_{2\mu}^{-1/2}}^\dag )$
with $T_3=-1/2$ ( figure 5(b) ).
The matrix element squared for all 4 diagrams contributing to the CC
u-channel process is
\bea
{ \left\vert{\cal M}_{LQ}^{u-chann}
(\nu_\mu {\bar u} \longrightarrow \mu^- {\bar b})\right\vert ^2 } =
\left[\hat u (\hat u - m_b ^2)\right]
\Biggl[ { \left\vert g_{1L} \, g_{1L} \right\vert ^2 \over
(\hat u - M_{S_1}^2)^2}
~+~
{ \left\vert g_{1L} \, g_{1R} \right\vert ^2 \over (\hat u - M_{S_1}^2)^2}
~+~
{ \left\vert g_{3L} \, g_{3L} \right\vert^2 \over (\hat u - M_{S_3^0}^2)^2}
\nonumber \\
~-~ 2 {\left\vert g_{1L} \, g_{3L}\right\vert^2 \over (\hat u - M_{S_1}^2) 
(\hat u - M_{S_{3}^0}^2) }\Biggr]
~+~ \left[ 4(\hat s + \hat u) (\hat s + \hat u - m_b ^2)\right]
\Biggl[{ \left\vert g_{2L}\,  g_{2L}
\right\vert^2 \over (\hat u - M_{ V _{2\mu} ^{-1/2}}^2)^2 } \Biggr]
\eea
\\

Next we consider the production of ``b'' from $\bar {\nu_e}$
(also obtained from the $\mu^-$ decay) through
interactions with nucleon via CC s-channel process
for $|F|=0$ case ( figure  6(a) )
and CC u-channel process for $|F|=2$ case  ( figure 6(b) ).
There are in all 4 diagrams contributing to production of b via
($\bar {\nu_e} \,+\, u \longrightarrow e^+ \,+\, b$)
in the s-channel ( figure 6(a) ),
one mediated by the charge $=2/3$,
scalar LQ $( R_2^{-1/2} )$ with $T_3=-1/2$ and
while the other three mediated by vector LQs
$( U_{1\mu}, U_{1\mu}, U_{3\mu}^0 )$ having $T_3=0$ each.
The matrix element squared for the 4 diagrams contributing to the
CC s-channel process is
\begin{figure}[h]
\vskip -4cm
\centerline{
\epsfxsize=20 cm\epsfysize=24cm\epsfbox{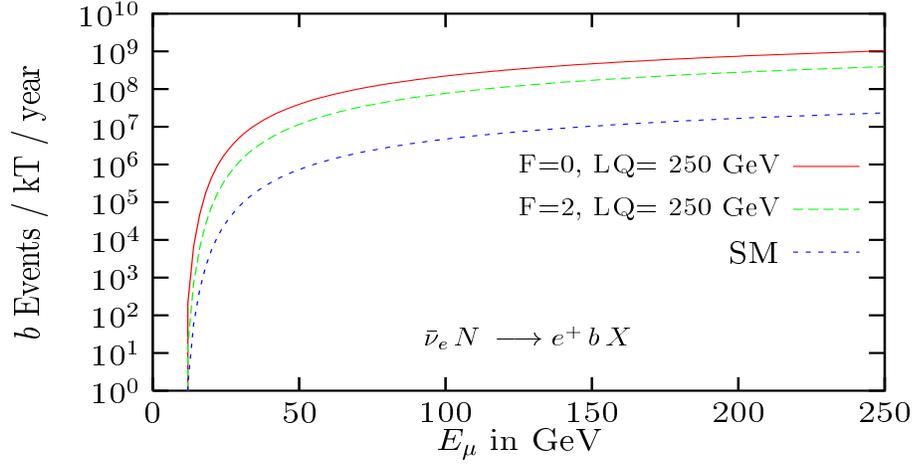}
}
\vskip -13cm

\caption{{\em Variation of $ b$-events ( from SM and LQ ) for a 1kT 
detector and
LQ mass 250 GeV with muon beam energy for a baseline length 40 meters
and sample detector area 0.025 m$^2$}}
\label{fig:fig8}
\end{figure}
\bea
{ \left\vert{\cal M}_{LQ}^{s-chann}(\bar {\nu_e} u
\longrightarrow e^+ b)\right\vert ^2 } =
\left[ \hat s (\hat s - m_b ^2)\right]
\Biggl[{ \left\vert {h_{2L}} \, {h_{2R}}
\right\vert ^2 \over (\hat s - M_{ {R _{2} ^{-1/2} }}^2)^2 }\Biggr]
~+~ \left[4 \hat u (\hat u - m_b ^2)\right]
\Biggl[
{\left\vert h_{1L}\, h_{1L} \right\vert ^2
\over (\hat s - M_{{U_{1\mu}}}^2)^2} \nonumber \\
~+~
{\left\vert h_{3L}\, h_{3L} \right\vert ^2
\over (\hat s - M_{{U_{3\mu}^0}}^2)^2}
~-~ 2 {\left\vert h_{1L}\, h_{3L} \right\vert ^2
\over (\hat s - M_{{U_{1\mu}}}^2) \,(\hat s - M_{{U_{3\mu}^0}}^2) }
\Biggr]
~+~
\left[4 (\hat s + \hat u)(\hat s + \hat u - m_b ^2)\right]
\Biggl[
{\left\vert h_{1L}\, h_{1R} \right\vert ^2
\over (\hat s - M_{{U_{1\mu}}}^2)^2} \Biggr]
\eea
\\
\noindent where, the Mandelstam variables at the parton level are given by
${\hat s}=(p_{\bar {\nu_e}}+p_u)^2$,
${\hat t}=(p_{\bar {\nu_e}}-p_{e^+})^2$ and
${\hat u}=(p_{\bar {\nu_e}}-p_b)^2$.

\par In the u-channel, three diagrams are mediated by the charge $=1/3$,
scalar LQs $( S_1, S_1, S_3^0 )$ with $T_3=0$ and
one is mediated by a vector LQ $( V_{2\mu}^{-1/2} )$ with $T_3=-1/2$
( figure 6(b) ).
The matrix element squared for all 4 diagrams contributing to the CC
u-channel process is
\bea
{ \left\vert{\cal M}_{LQ}^{u-chann}(\bar {\nu_e} u \longrightarrow e^+ b)\right\vert ^2 } =
\left[\hat u (\hat u - m_b ^2)\right]
\Biggl[ { \left\vert g_{1L} \, g_{1L} \right\vert ^2
\over (\hat u - M_{S_1}^2)^2}
~+~
{ \left\vert g_{1L} \, g_{1R} \right\vert ^2
\over (\hat u - M_{S_1}^2)^2}
~+~
{ \left\vert g_{3L} \, g_{3L} \right\vert^2
\over (\hat u - M_{S_3^0}^2)^2} 
\nonumber \\
~-~ 2
{\left\vert g_{1L} \, g_{3L}\right\vert^2
\over (\hat u - M_{S_1}^2)  (\hat u - M_{S_{3}^0}^2) }\Biggr]
~+~ \left[ 4(\hat s + \hat u) (\hat s + \hat u - m_b ^2)\right]
\Biggl[{ \left\vert g_{2L}\,  g_{2L}\right\vert^2
\over (\hat u - M_{ V _{2\mu} ^{-1/2}}^2)^2 } \Biggr]
\eea
\\

\begin{figure}[h]
\vskip -4cm
\centerline{
\epsfxsize=20 cm\epsfysize=24cm\epsfbox{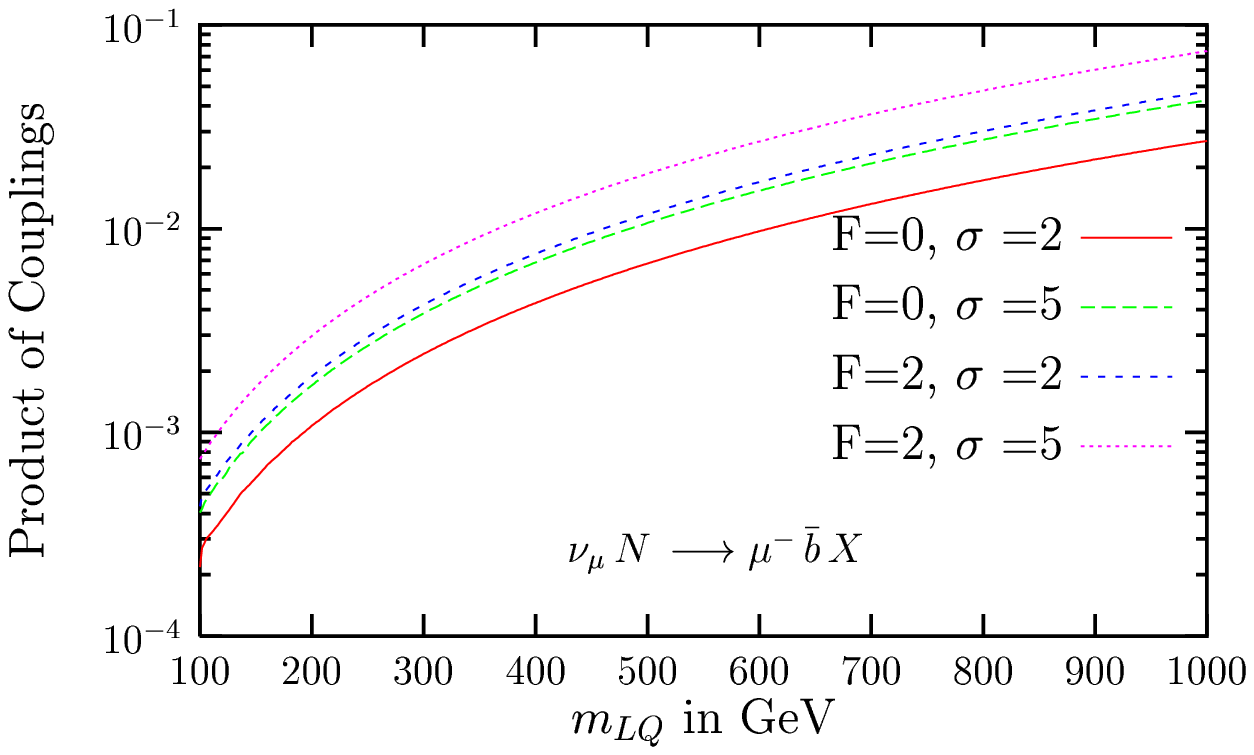}
}
\vskip -13cm
\caption{{\em Contour plot for $\bar b$ production  at
2$\sigma$ and 5$\sigma$ effect for $E_\mu$=50 GeV,
baseline length=40 meters and
sample detector of area 2500 cm$^2$ and mass 1kT.}}
\label{fig:fig9}
\end{figure}

\begin{figure}[h]
\vskip -4cm
\centerline{
\epsfxsize=20 cm\epsfysize=24cm\epsfbox{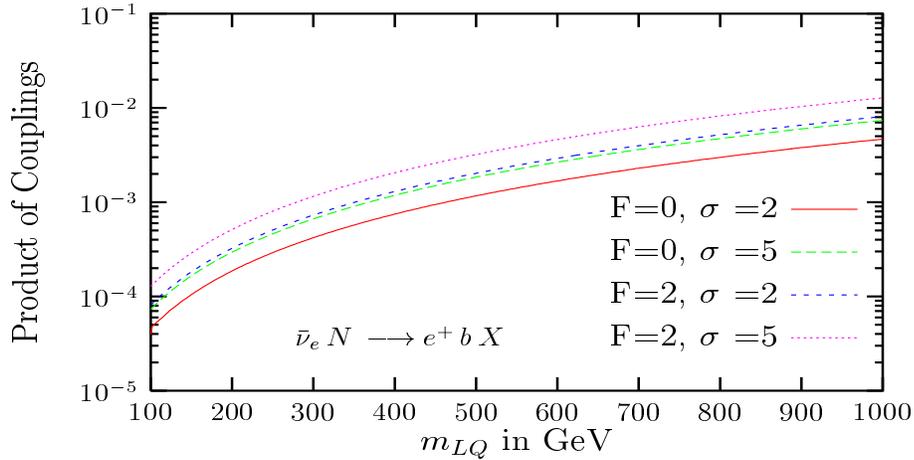}
}
\vskip -13cm
\caption{{\em Contour plot for $ b$ production  at
2$\sigma$ and 5$\sigma$ effect for $E_\mu$=50 GeV,
baseline length=40 meters and
sample detector of area 2500 cm$^2$ and mass 1kT.}}
\label{fig:fig10}
\end{figure}

\par As discussed in the previous section,
we have used the model independent bounds on couplings
from \cite{davidson} and the relevant bounds for the processes listed above
are listed in table \ref{table1}.
We extract bounds relevant to $(\nu_\mu \bar u)(\mu^- \bar b)$ vertex
from the bounds for (21)(23) generation of quark-lepton pair, while for the
vertex $(\bar \nu_e u)(e^+ b)$, we use the bounds for
the (11)(13) generation indices relevant for the process $
\bar \nu_e u\longrightarrow e^+ b$.
The other inputs to compute the event rates are the same as
for the NC diagrams.
In figures 7 and 8, we have plotted the $\bar b$ and $b$ event rates as a function of muon beam energy 
for $\nu_\mu-N$ and ${\bar \nu}_e-N$ scattering processes respectively. For these processes 
we have also plotted the SM contribution to $\bar b$ and b events. To determine the 
allowed range of LQ masses and products of couplings, we have used the criterion that the 
number of signal events is equal to two or five times the square root of events in the SM. 
Accepting this requirement of 2$\sigma$ and 5$\sigma$ effect as a sensible discovery criterion 
the contour in figures 9 and 10 are drawn for a baseline length of 40 m and thus the non-compliance 
of these estimates with experimental observation would mean that the region above these 
curves is ruled out. 
\end{section}  
\begin{section}{Conclusions and Discussion}
Heavy quark $(b, \bar b)$ production from $\nu_\mu-{\rm N}$ and 
${\bar \nu}_e-N$ scattering via both the CC and NC interactions at a 
NF provides an exciting possibility to detect signals of new physics. 
This comes about because in these processes the SM contribution is heavily  
suppressed either due to CKM matrix element or due to interaction of 
neutrinos with the sea quarks present inside the nucleon. 
The NC processes in SM are further suppressed as they can 
take place only at one loop level. 
We have computed here the $b(\bar b)$ event rates in theories with LQ and 
confined ourselves to the near-site experiments where the 
oscillation effects are negligible. 
From figure 7, it is clear that 
the contribution coming from the SM 
to the $\bar b$ production rate in the CC channel is higher than that of
 LQ's with $\vert {\rm F} \vert=0$, while it is lower than the 
contribution from LQ's with $\vert {\rm F} \vert=2$ 
for our choice of the couplings obtained from low energy experiments.
We typically find (figure 8) that the SM contribution to $b$ 
production rate is 2 to 3 orders of magnitude smaller than LQ contribution 
in CC channel even after using the most severe constraints on 
LQ couplings and masses from low energy FCNC processes. 
Further the $b$ production rate in the NC channel (figures 3 and 4) 
is comparable to that for the CC case.
We have investigated the region in coupling~-~mass 
space for LQ which can provide a reasonable signal for the 
discovery of new physics involving LQ. It may be noted that this region 
can be even more restrictive than that implied by the low energy 
bounds obtained from B meson decays.  
Also the inclusion of LFV interactions via LQ's 
could further squeeze the allowed region of 
LFV Couplings and masses.

\end{section}
\noindent{\bf Acknowledgment :} P.M. acknowledges 
Council for Scientific and Industrial Research,
India while A.G. acknowledges the University
Grants Commission, India for partial financial support.
We also thank SERC, Department of Science \& Technology, New Delhi,  for
partial financial support.
\newcommand{\plb}[3]{{Phys. Lett.} {\bf B#1,} #2 (#3)}                  %
\newcommand{\prl}[3]{Phys. Rev. Lett. {\bf #1,} #2 (#3)}        %
\newcommand{\rmp}[3]{Rev. Mod.  Phys. {\bf #1,} #2 (#3)}             %
\newcommand{\prep}[3]{Phys. Rep. {\bf #1,} #2 (#3)}                     %
\newcommand{\rpp}[3]{Rep. Prog. Phys. {\bf #1,} #2 (#3)}             %
\newcommand{\prd}[3]{Phys. Rev. {\bf D#1,} #2 (#3)}                    %
\newcommand{\prc}[3]{{Phys. Rev.}{\bf C#1,} #2 (#3)} 
\newcommand{\np}[3]{Nucl. Phys. {\bf B#1,} #2 (#3)}                    %
\newcommand{\npbps}[3]{Nucl. Phys. B (Proc. Suppl.)
          {\bf #1,} #2 (#3)}                                           %
\newcommand{\sci}[3]{Science {\bf #1,} (#3) #2}                 %
\newcommand{\zp}[3]{Z.~Phys. C{\bf#1,} #2 (#3)}                          
\newcommand{\ijmpa}[3]{Int.~J. Mod. Phys.{\bf A#1,} #2 (#3)}     
\newcommand{\ijmpasu}[4]{Int.~J. Mod. Phys.{\bf A#1,} Proc. Suppl. 
#2, #3 (#4)}     
\newcommand{\mpla}[3]{Mod. Phys. Lett. {\bf A#1,} #2 (#3)}             %
\newcommand{\epjc}[3]{Euro. Phys. J.{\bf C#1,} #2 (#3)}
\newcommand{\apj}[3]{ Astrophys. J.\/ {\bf #1,} #2 (#3)}       %
\newcommand{\astropp}[3]{Astropart. Phys. {\bf #1,} #2 (#3)}            %
\newcommand{\ib}[3]{{ ibid.\/} {\bf #1,} #2 (#3)}                    %
\newcommand{\nat}[3]{Nature (London) {\bf #1,} (#3) #2}         %
\newcommand{\app}[3]{{ Acta Phys. Polon.   B\/}{\bf #1,} (#3) #2}%
\newcommand{\nuovocim}[3]{Nuovo Cim. {\bf C#1,} (#3) #2}         %
\newcommand{\yadfiz}[4]{Yad. Fiz. {\bf #1,} (#3) #2;             %
Sov. J. Nucl.  Phys. {\bf #1,} #3 (#4)]}               %
\newcommand{\jetp}[6]{{Zh. Eksp. Teor. Fiz.\/} {\bf #1,} (#3) #2;
          {JETP } {\bf #4} (#6) #5}%

\newcommand{\philt}[3]{Phil. Trans. Roy. Soc. London A {\bf #1} #2 
  (#3)}                                                          %
\newcommand{\hepph}[1]{(hep--ph/#1)}           %
\newcommand{\hepex}[1]{(hep--ex/#1)}           %
\newcommand{\astro}[1]{(astro--ph/#1)}         %


\begin{thebibliography}{99}

\bibitem{storage_ring}
S.~Dutta, R.~Gandhi, B.~Mukhopadhyaya,  
\epjc{18}{405}{2000};
C.~Quigg, \hepph{9803326};
S.~Geer, \prd{57}{6989}{1998};   
D.~Ayres {\it et al.}, (electronic archive: physics/9911009);
A.~Cervera {\it et al.}, \hepph{0002108};
A.~Blondel {\it et al.}, CERN-EP-2000-05;
C. Albright {\it et al.}, \hepex{0008064};  
S.~Geer, \hepph{0008155}.
\bibitem{hepph010} M.L. Mangano {\it et. al.}, \hepph{0105155}.
\bibitem{fukuda} Y.~Fukuda et. al., \plb{433}{9}{1998},
\prl{81}{1562}{1998}; T.~Kagita, in proceedings of the XVIIIth
International Conference on Neutrino Physics and Astrophysics,Takayama,
Japan (June 1998).

\bibitem{solarmsw} L.~Wolfenstein, \prd{17}{2369}{1978},
\prd{20}{2634}{1979}; S.~P.~Mikheyev and A.~Yu~Smirnov, Sov. J.
Nucl. Phys.{\bf 42} (1986) 913.
  
\bibitem{solarvac} B.~Pontecorvo, Sov. Phys. JETP {\bf 26} (1968) 984.

\bibitem{rp}
A.~Datta, R.~Gandhi, B.~Mukhopadhayaya, P.~Mehta, \prd{64}{015011}{2001}.

\bibitem{lq}
P.~Mehta, A.~Goyal, S.~Dutta, \plb{535}{219}{2002}.

\bibitem{heavy}
D.~Chakraverty, A.~Dutta, B.~Mukhopadhayaya, \plb{503}{74}{2001}.

\bibitem{lqs}
J.~C.~Pati and Abdus Salam, \prd{10}{275}{1974};
O.~Shankar, \np{206}{253}{1982};
W.~Buchmuller and D.~Wyler, \plb{177}{377}{1986};
W.~Buchmuller, R.~Ruckl, D.~Wyler, \plb{191}{442}{1987};
P.~Langacker, M.~Luo, Alfred~K~Mann, \rmp{64}{87}{1992};
J.~Blumlein and R.~Ruckl, \plb{304}{337}{1993};
M.~A.~Doncheski and R.~W.~Robinett, \prd{56}{7412}{1997};
U.~Mahanta, \prd{62}{073009}{2000}.

\bibitem{lq_lag} See the third and sixth references in \cite{lqs}.

\bibitem{davidson}
S.~Davidson, D.~Bailey, B.~A.~Campbell, \zp{61}{613}{1994};
E.~Gabrielli, \prd{62}{055009}{2000};

\bibitem{lqbounds}S.~Aid {\em et al.} \plb{353}{578}{1995};
Derrick {\em et al.}, \zp{73}{613}{1997};
C.~Adloff {\em et al.}, \hepex{9907002}.

\bibitem{cteq}H.~Lai {\it et al.}, \prd{55}{1280}{1997}.

\bibitem{detector} C.~Albright {\it et al.}, \hepex{0008064}.

\bibitem{baba_belle} K.~Abe {\it et. al. } (The Belle Collaboration), 
\hepex{0204002}, to appear in Phys. Rev. Lett.;  
B.~Aubert {\it et. al. } (BABAR Collaboration), \hepex{0207080};  
A.~Roodman, \hepex{0112019}; 
V.~Halyo, \hepex{0207010};  
J.~Nam {\it et. al.} (The Belle Collaboration), \ijmpasu{16}{1B}{625}{2001}.
\end{thebibliography}
\end{document}